\documentclass[aps,prb,showpacs]{revtex4}
\usepackage{graphicx,amsmath,amssymb,color}
\usepackage{float}
\begin{document}
\title{Non-local Plasma Spectrum of Graphene interacting with a Thick Conductor}
\author{Godfrey Gumbs$^{1,2}$,  Andrii Iurov$^{1,4}$,  and N. J. M. Horing$^3$}
\affiliation{$^{1}$Department of Physics and Astronomy, Hunter College of the
City University of New York, 695 Park Avenue, New York, NY 10065, USA\\
$^{2}$Donostia International Physics Center (DIPC),
P de Manuel Lardizabal, 4, 20018 San Sebastian, Basque Country, Spain\\
$^{3}$Department of Physics and Engineering Physics, Stevens Institute of Technology, Hoboken, NJ 07038, USA \\
$^{4}$Center for High Technology Materials, University of New Mexico, NM 87106, USA}

\date{\today}

\begin{abstract}

Self-consistent field theory is used to obtain the non-local plasmon dispersion
relation of monolayer graphene which is Coulomb-coupled to a thick
conductor. We calculate numerically the undamped plasmon excitation
spectrum for arbitrary wave number. For gapped graphene, both  the
low-frequency (acoustic) and high frequency (surface) plasmons
may lie within an undamped opening in  the particle-hole region.
Furthermore, we obtain plasmon excitations in a region of
frequency-wave vector space which do not exist for free-
standing gapped graphene.

\end{abstract}

\vskip 0.2in


\maketitle\section{Introduction}
\label{sec1}

Recent research on plasmon excitations
\cite{RMP, chakraborty, CambridgeBook, GumbsBook} has covered fundamental
aspects such as nonlocality \cite{DH},
quantum effects in nanoscale structures  including
fullerene \cite{our1, our2, Anto}, graphene \cite{Wunsch, pavlo}, carbon
nanotubes \cite{nanotubes1Lin, nanotubes2}, silicene  \cite{Silicene, silicene-canada}
and metallic dimers \cite{Nord1},
surface plasmon lasing \cite{Stockman}, plasmon-electron interaction \cite{Tege}
 and the potential role played by   plasmon
excitations in electronic sensors  \cite{sens1, Arx2} and  radiation
degradation of electronic and optoelectronic devices \cite{rad}.
The surge in activity to understand and discover  novel plasmonic materials
is stimulated by possible  applications  such as light concentration
for solar energy \cite{solar},  devices for telecommunications \cite{telecom},
and near-field instrumentation \cite{near}. Investigation of the damped terahertz 
plasmons in grpahene, interacting with sruface plasmons of a substrate with a 
large doping due to a large scattering rate, was addressed in \cite{satou}. 
The authors demonstrated that the field spread of the graphene plasmons into
the substrate suppressed.

\medskip
\par

In view of the stated importance of achieving a detailed understanding of plasma excitations,
we devote this paper to  a specific area which has not been adequately covered
so far in the literature.  It concerns plasmon excitations in monolayer graphene.
There are several papers dealing with calculations of the dispersion relation for
monolayer graphene that is doped \cite{Wunsch, 7+, gumbs, silkin} as well as pristine
graphene whose collective charge density oscillations  are driven by temperature
\cite{DasSarma}. The work on gapped graphene \cite{pavlo}  was partially
motivated by the observation that when monolayer graphene is on a substrate such  as
boron nitride, an energy gap between the valence and conduction
bands is produced yielding a plasmon and single-particle excitation spectrum which
can be drastically different from gapless monolayer graphene.  In Refs.
\cite{Wunsch} through \cite{pavlo}, a detailed  calculation of the undamped
plasmon excitations was carried out for all wavelengths.  Although computationally
challenging,   these calculations proved useful since our goal is to obtain
a full understanding of the response properties of nanoscale structures to
external probes. In a recent  paper
\cite{1}, it was demonstrated that the plasmon excitations in graphene has a linear
dispersion rather than a square root dependence on the wave vector. This startling result
came as a surprise because  since theoretical calculations on free-standing graphene clearly
does not yield a  linear dependence in the long wavelength limit.
As a matter of fact, this linear dependence of plasmon frequency  on wave vector was initially attributed
to local field corrections to the random-phase approximation. Horing \cite{NJMH}
showed that when graphene is Coulomb-coupled to a conductor, the surface plasmon
causes the low-frequency $\pi$-plasmon to  have a linear dispersion. In this paper, we
calculate the full dispersion relation for undamped plasmons in a hybrid
monolayer graphene-conductor structure. We exploit our simulations to consider
how the plasmon dispersion is affected when there is an energy gap between the
valence and conduction bands, thereby generalizing the results in \cite{pavlo}
where a surface is assumed to play a role.

The longitudinal  excitation spectra
of allowable modes will be determined from a knowledge
of the frequency-dependent non-local dielectric function
$ \epsilon ({\bf r},{\bf r}^\prime;\omega)$  of the composite system,  which depends
on the position coordinates ${\bf r},\  {\bf r}^\prime$ and
frequency $\omega$. Alternatively, the normal modes correspond to the resonances
of the inverse  dielectric function
$ K({\bf r},{\bf r}^\prime;\omega)$ satisfying
$\int d{\bf r}^\prime\  K({\bf r},{\bf r}^\prime;\omega) \epsilon({\bf r}^\prime,{\bf r}^{\prime\prime};\omega)
=\delta({\bf r}- {\bf r}^{\prime \prime})$.   The significance of
$K({\bf r},{\bf r}^\prime;\omega)$ is that it embodies many-body effects \cite{AI2,Horing} through
screening by the medium of an external potential $U({\bf r}^\prime;\omega)$ to
produce an effective potential $V({\bf r};\omega)=\int d{\bf r}^\prime\ K({\bf r},{\bf r}^\prime;\omega)
U({\bf r}^\prime;\omega)$.  In Sec.\ \ref{sec2}, we briefly review the formalism
for calculating the inverse dielectric function for a 2D layer interacting
with a semi-infinite conductor. Section \ref{sec3} is devoted to
our numerical results for the dispersion relations at arbitrary
wavelength  for this hybrid structure. We show explicitly  how
the gap for monolayer graphene  affects both the dispersion relation
for the surface plasmon and the low-frequency acoustic mode. Specifically,
we demonstrate that  the low-frequency plasmon branch may exist
in a region of frequency-wave vector space that was not obtained
for free standing gapped graphene. We conclude with a summary of
our results in Sec.\ \ref{sec4}.

\section{General Formulation of the Problem}
\label{sec2}

\begin{figure}[H]
\centering
\includegraphics[width=0.35\textwidth]{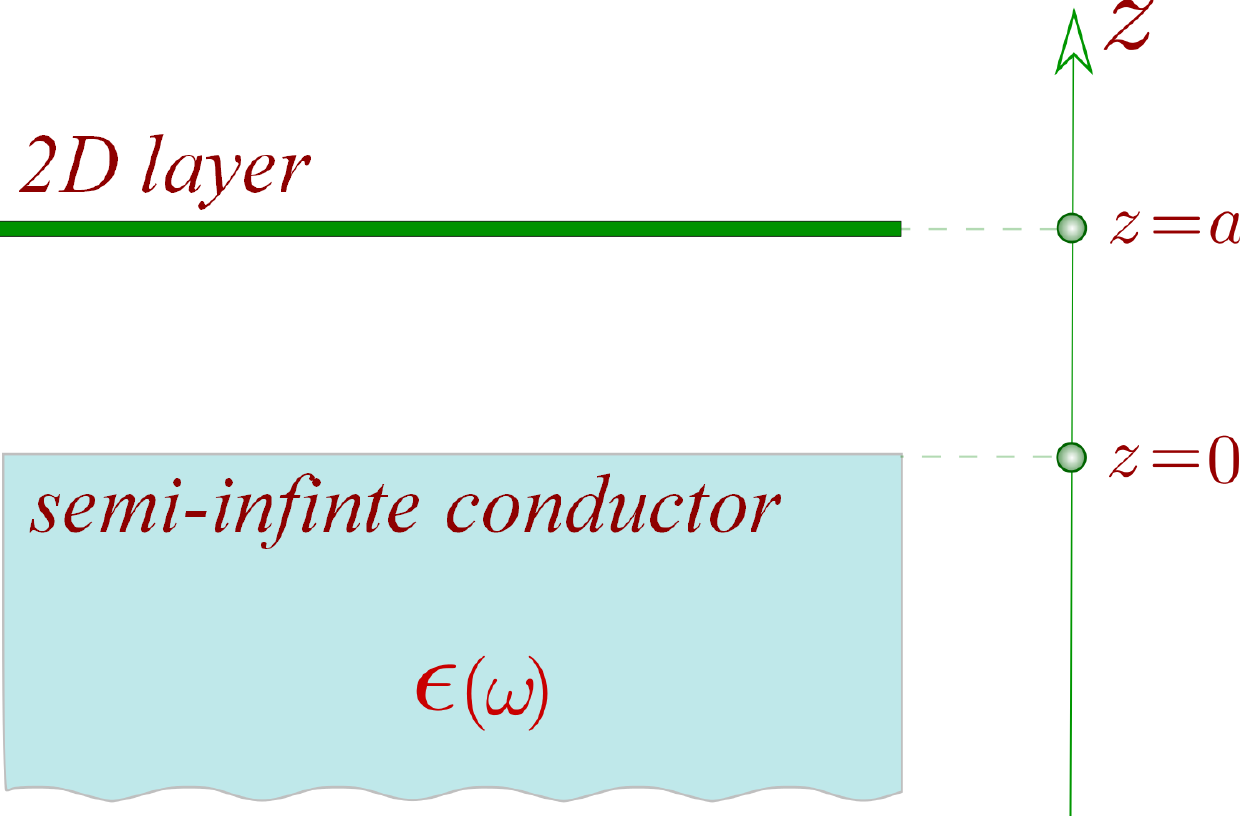}
\caption{(Color online) Schematic illustration of a thick (semi-
infinite)  metallic plasma interacting through the Coulomb force with a thin layer such as monolayer graphene.}
\label{FIG:1}
\end{figure}

In   this work, we consider a  composite nano-scale system consisting
of a   2D layer  separated from a thick dielectric material.  The 2D
layer may be monolayer graphene (or a 2DEG such as a
semiconductor  inversion layer or HEMT (high electron mobility
transistor)).  The 2D graphene  layer may have a gap, thereby
broadening the applicability  of the composite system model which
also incorporates a  separation layer and a semi-infinite  plasma,  as
depicted in Fig.\ \ref{FIG:1}. The excitation spectra
of allowable modes will be determined from a knowledge
of the non-local dielectric function
$ \epsilon ({\bf r},{\bf r}^\prime;\omega)$   which depends
on position coordinates ${\bf r}, {\bf r}^\prime$ and
frequency $\omega$ or its inverse
$ K({\bf r},{\bf r}^\prime;\omega)$ satisfying
$\int d{\bf r}^\prime\  K({\bf r},{\bf r}^\prime;\omega) \epsilon({\bf r}^\prime,{\bf r}^{\prime\prime};\omega)
=\delta({\bf r},{\bf r}^{\prime \prime})$.  The self-consistent
field structure for $K({\bf r},{\bf r}^\prime;\omega)$ is  determined, using the
technique of Ref. \ [\onlinecite{NJMH}].

In operator notation, the  composite dielectric function  $\hat{\epsilon}$ and its inverse,
$ \hat{K}= \hat{\epsilon}^{-1} $, for the 2D layer and semi-infinite
substrate  is given by adding their polarizabilities  $\hat{\alpha}_{2D}$ and
 $\hat{\alpha}_{SI}$, respectively, i.e.,

\begin{equation}
\hat{K}^{-1}=\hat{\epsilon}= \hat{1}+\hat{\alpha}_{SI}+\hat{\alpha}_{2D} \equiv \hat{\epsilon}_{SI}+ \hat{\alpha}_{2D}
=\hat{K}_{SI}^{-1}+  \hat{\alpha}_{2D} \ .
\label{one}
\end{equation}
Multiplication of Eq.\ (\ref{one})  from the right by $\hat{K}$ and left by $\hat{K}_{SI}$
yields the basic random-phase approximation (RPA) integral
equation

\begin{equation}
\hat{K}=\hat{K}_{SI} -\hat{K}_{SI} \cdot  \hat{\alpha}_{2D}\cdot  \hat{K} \ .
\end{equation}
 Additionally, $\hat{K}_{SI}$ is the
inverse dielectric function for the semi-infinite substrate alone,
whose surface lies in the $z=0$ plane.
In  explicit integral form, after Fourier transforming with respect to coordinates
 parallel to the translationally invariant $xy$-plane and
suppressing the in-plane wave number $q_{||}$ and frequency $\omega$,
we obtain

\begin{figure}[ht!]
\begin{centering}
\includegraphics[width=0.51\textwidth]{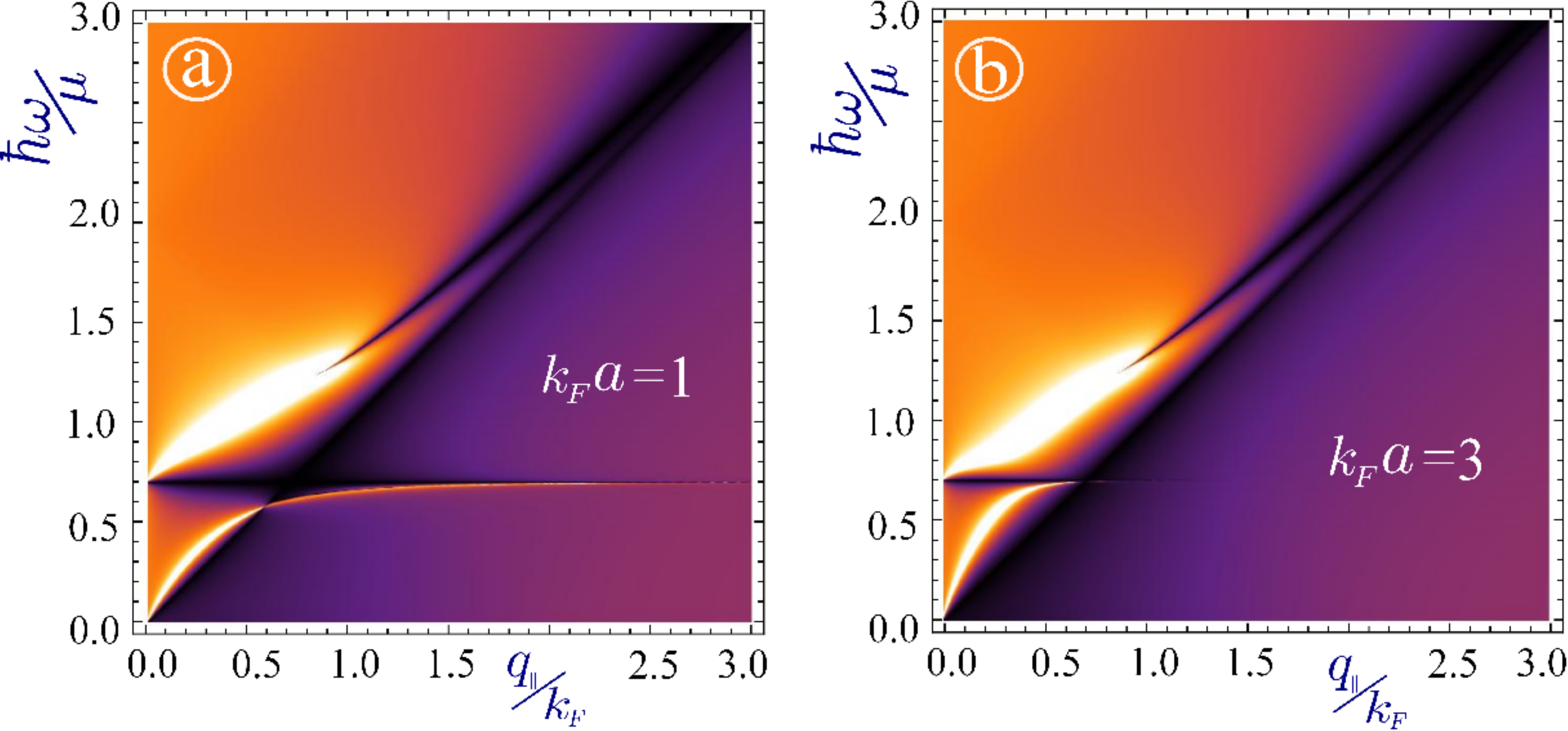}
\includegraphics[width=0.51\textwidth]{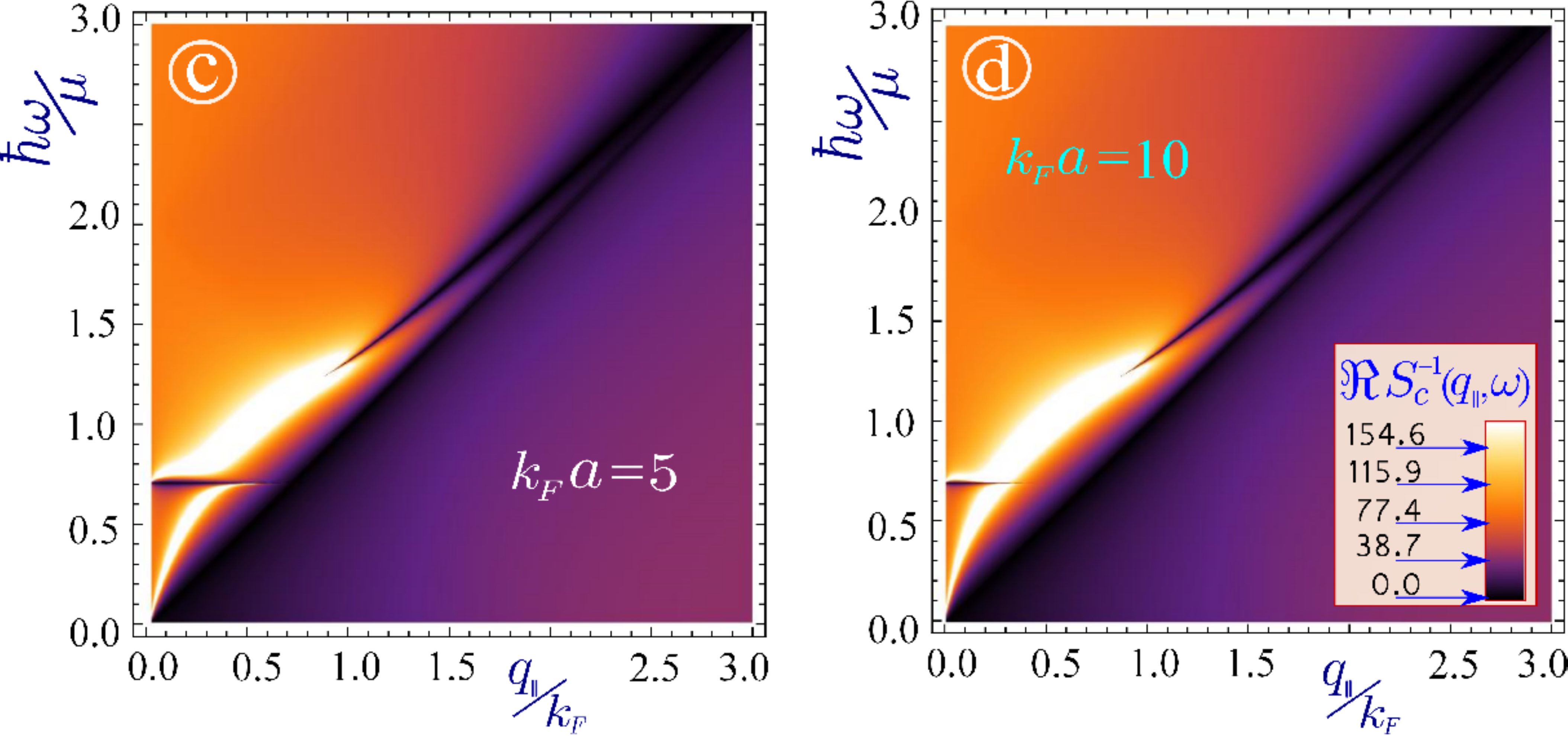}
\end{centering}
\caption{(Color online)  Density plots of the real part of the inverse of the dispersion function
$S_c(q_{\parallel}, \omega+i0^+)$  for extrinsic 
(doped) graphene with no band gap $(\Delta = 0)$  with the peaks  corresponding to the plasmon 
resonances.   Panels $(a)-(d)$ demonstrate the
 plasmon spectrum for various separations between the graphene layer
and the surface - $a = 1, \, 3, \, 5 \, \text{and} \, 10 \, k_F^{-1}$, where $k_F $ is the Fermi wave number.}
\label{FIG:2}
\end{figure}

\begin{figure}[ht!]
\begin{centering}
\includegraphics[width=0.51\textwidth]{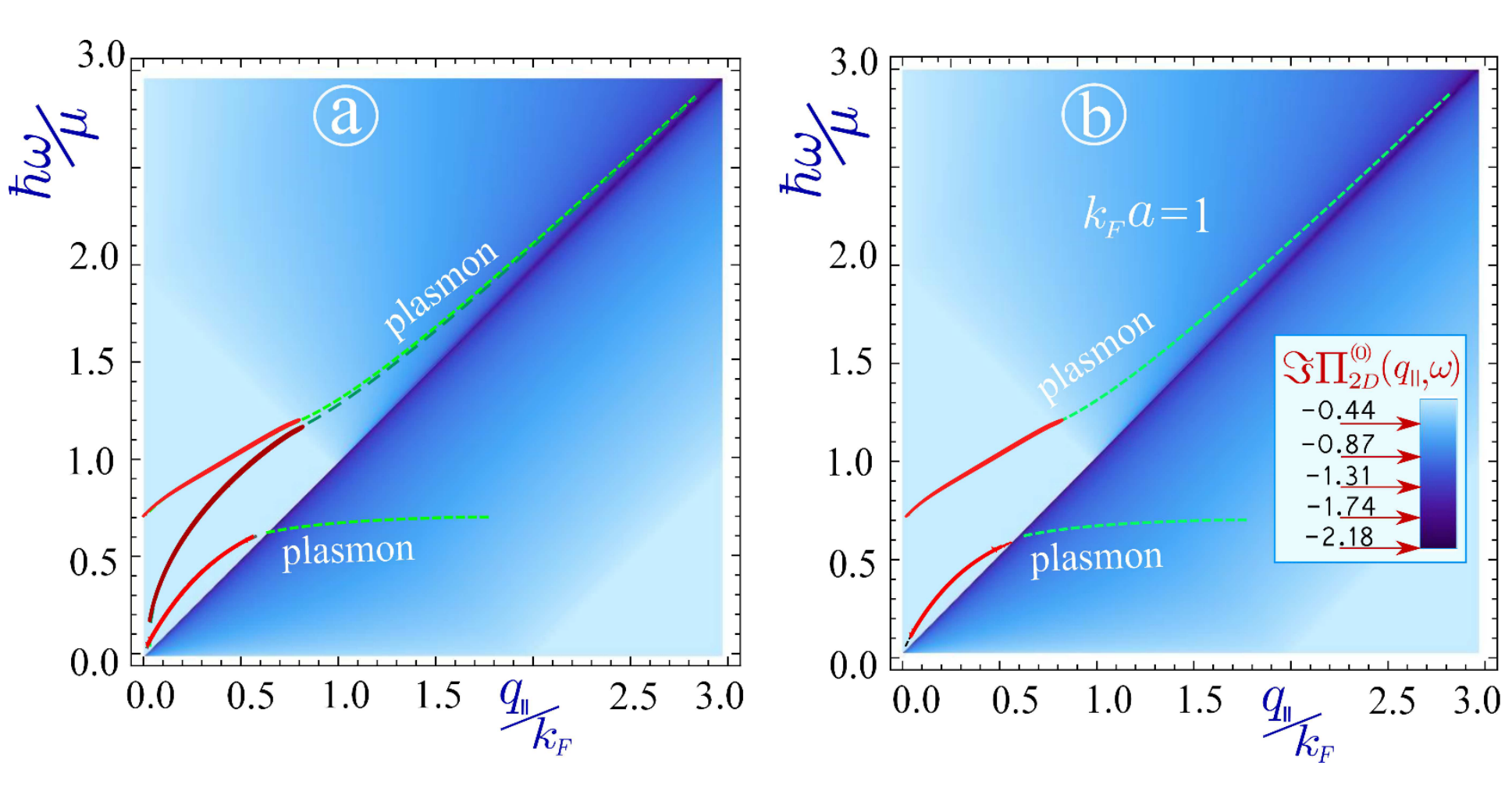}
\end{centering}
 \caption{(Color online)  Exact numerical solutions for the plasmon dispersion
of gapless graphene.  In (a), the highest and lowest curves are the solutions
 of the plasmon  dispersion equation $\Re S_C(q_\parallel,\omega)=0$  for graphene
at a distance  $a=k_F^{-1}$ from a  conducting  surface,
  whereas the curve in between these two solutions corresponds to the zeros of
$  1+ 2\pi e^2/(\epsilon_s q_\parallel) \Re \Pi_{2D}^{(0)} (q_{||},\omega)=0  $
 for free standing graphene.  In (b), only the solutions for $\Re S_C(q_\parallel,\omega)=0$
are presented.  In both (a) and (b),  the plasmon energy is scaled with respect to the
chemical potential $\mu$ and  we superimpose all plasmon curves on a background of a density plot
of $\Im\ \Pi_{2D}^{(0)} (q_{||},\omega)$ to illustrate the effects due to Landau damping.
  }
\label{FIG:3}
\end{figure}

\begin{figure}[ht!]
\centering
\includegraphics[width=0.51\textwidth]{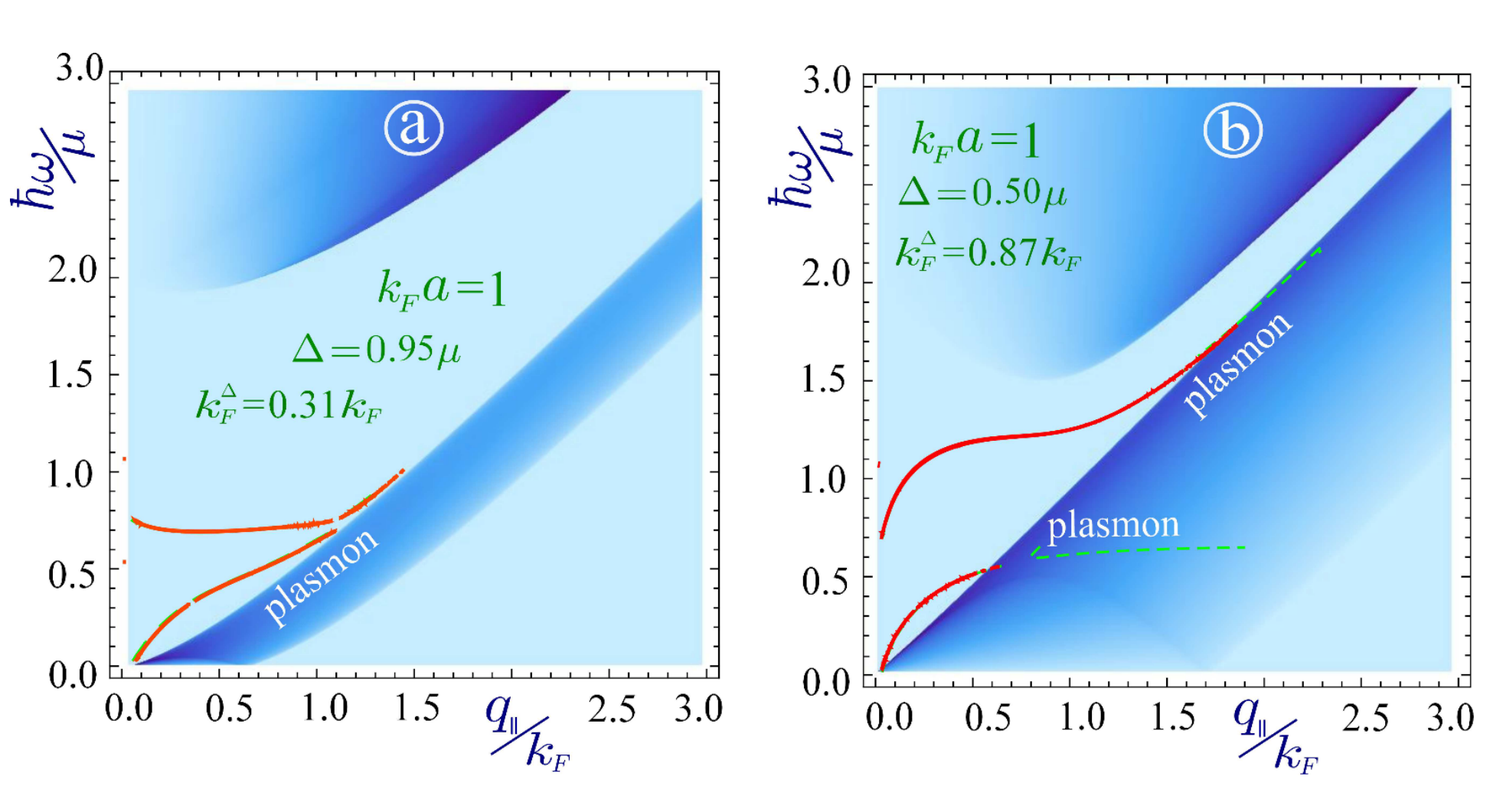}
\includegraphics[width=0.51\textwidth]{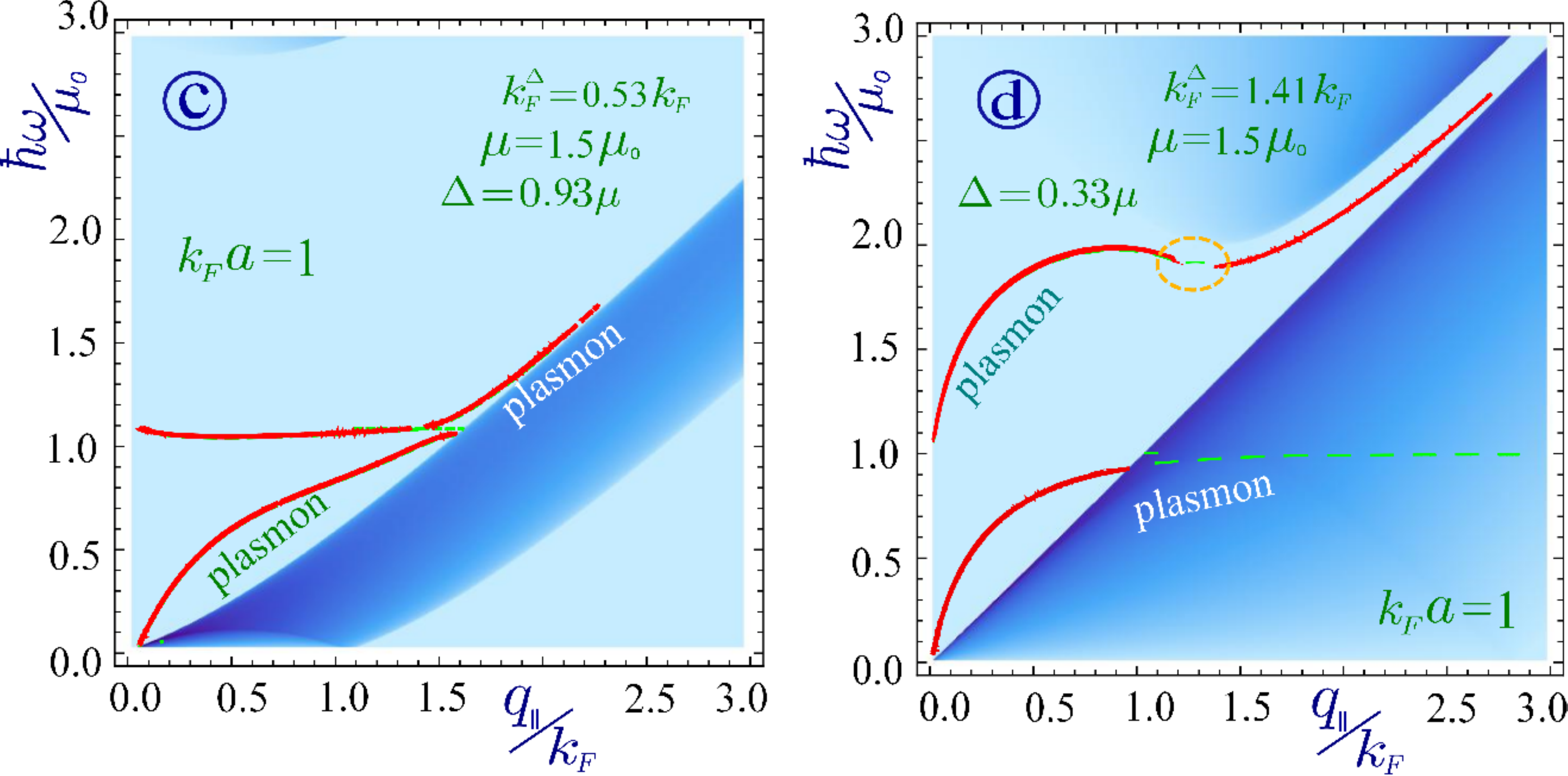}
\caption{(Color online)  Exact numerical solutions of $\Re\ S_C(q_\parallel,\omega+i0^+)=0$
for gapped graphene  at a distance $a=k_F^{-1}$  from a conducting surface.
The plasmon  excitation spectrum is superimposed on a background showing the density plot of
$\Im \ \Pi_{2D}^{(0)}(q_\parallel,\omega+i0^+)$ whose values determine Landau damping.
The red lines  correspond to undamped plasmons when the magnitude of the 
plasmon dispersion function  $\left |  S_C \left( q_\parallel,\omega+i0^+\right)   \right|$ vanishes.
 	Panels $(a)$ and $(b) $ show the case of
	$\Delta = 0.95 \, \text{and} \, 0.5$, panels $(c)$ and $(d)$ demonstrate the behavior of
	the 	plasmon spectra for $\mu = 1.5 \mu_0$ and $\Delta = 0.93 \mu$ and $\Delta =
	0.33 \mu$, respectively.  Here, $\mu_0=0.2$ eV is the chemical potential used in the calculations
	of Fig.\ \ref{FIG:3}. 	This value for $\mu_0$ is chosen to ensure the applicability 
	of isotropy of the energy band structure at low doping.\cite{silkin} 
	Also, in our notation, $k_F^\Delta\equiv \sqrt{\mu^2-\Delta^2}/(\hbar v_F)$.}
\label{FIG:4}
\end{figure}

\begin{equation}
K(z_1,z_2)= K_{SI}(z_1,z_2) - \int_{-\infty}^\infty dz^\prime \int_{-\infty}^\infty
 dz^{\prime\prime}\  K_{SI} (z_1,z^\prime)
\alpha_{2D}(z^\prime ,z^{\prime\prime})  K(z^{\prime\prime} ,z_2) \ .
\end{equation}
Here, the polarization function for the 2D layer is given by

\begin{equation}
\alpha_{2D}(z^{\prime}, z^{\prime\prime})= \int_{-\infty}^\infty  dz^{\prime\prime\prime} \
v(z^\prime, z^{\prime\prime\prime} ) D(z^{\prime\prime\prime},z^{\prime\prime}) \ ,
\end{equation}
where $v$ is the Coulomb potential energy and the 2D response function's localization to
the layer at $z=a$ is expressed as

\begin{equation}
D(z^{\prime\prime\prime},z^{\prime\prime}) = \Pi_{2D}^{(0)} (q_{||},\omega)
  \delta(z^{\prime\prime\prime}-a) \delta(z^{\prime\prime}-a) \ , 
\end{equation}
with $\Pi_{2D}^{(0)} (q_{||},\omega)$  as the 2D ring diagram.  Upon
substituting this form of the polarization function for the monolayer into
the integral equation for the composite inverse dielectric function $K$, we have

\begin{equation}
K(z_1,z_2)= K_{SI}(z_1,z_2) -
\Pi_{2D}^{(0)} (q_{||},\omega) \int_{-\infty}^\infty  dz^\prime\ K_{SI}(z_1,z^\prime) v(z^\prime-a) K(a,z_2)\ .
\label{eq:GG}
\end{equation}
We now set $z_1=a$ in Eq.\ (\ref{eq:GG}) and obtain

\begin{equation}
K(a,z_2)= K_{SI}(a,z_2) -
\Pi_{2D}^{(0)} (q_{||},\omega) \left\{\int_{-\infty}^\infty  dz^\prime\ K_{SI}(a,z^\prime) v(z^\prime-a)\right\}
 K(a,z_2)\ .
\label{eq:GG2}
\end{equation}
Solving algebraically for $K(a,z_2)$ yields

\begin{equation}
K(a,z_2) = \frac{ K_{SI}(a,z_2)}{S_{C}(q_{||},\omega)}
\label{eq:GG3}
\end{equation}
with the ``dispersion function"'       $S_{C}(q_{||},\omega) $ given by

\begin{equation}
S_{C}(q_{||},\omega) \equiv 1+ \Pi_{2D}^{(0)} (q_{||},\omega) \left\{\int_{-\infty}^\infty  dz^\prime\
 K_{SI}(a,z^\prime) v(z^\prime-a)\right\} \ ,
\label{nine}
\end{equation}
whose zeros determine the plasmon resonances of the composite system. In our numerical calculations, we
employ
$K_{SI} (z,z^\prime)$ given by Eq. (30) of Ref. \  [\onlinecite{Horing}] for the semi-infinite  metallic substrate
in the local limit, whence Eqs. (\ref{eq:GG}) through  (\ref{nine})  yield \cite{NJMH}

\begin{equation}
K(z_1,z_2)= K_{SI}(z_1,z_2) -
\Pi_{2D}^{(0)} (q_{||},\omega)\frac{ K_{SI}(a,z_2)}{S_{C}(q_{||},\omega)}
\left\{ \int_{-\infty}^\infty \  dz^\prime\ K_{SI}(z_1,z^\prime) v(z^\prime-a) \right\}  \
\label{eq:GG-inv}
\end{equation}
with

\begin{equation}
S_{C}(q_{||},\omega)= 1+\frac{2\pi e^2}{\epsilon_s q_{||}} \Pi_{2D}^{(0)} (q_{||},\omega)
\left\{ 1+ e^{-2q_{||} a}  \frac{1-\epsilon_B(\omega)}{1+\epsilon_B(\omega)}   \right\} \ .
\label{eleven}
\end{equation}

Although the principal focus here is to examine the role of 2D graphene plasma nonlocality
embedded in $\Pi^{(0)}_{2D}(q_\parallel,\omega)$  on the coupled  plasmon spectrum of the
composite  system,  we briefly revisit the local results of Ref.[\onlinecite{NJMH}] to point out
their generalization to include gapped graphene  along with the previously discussed
gapless results.  In this regard, the graphene polarizability is also taken in the local limit with
$ \Pi_{2D}^{(0)} (q_{||},\omega)\approx Cq_\parallel^2/\omega^2$ so that
	Eq. \ (\ref{eleven}) yields

 \begin{equation}
1-\frac{2 \pi  C e^2}{\epsilon_s \omega^2}q_{\parallel} \left\{
1+  e^{-2 a q_{\parallel}} \frac{\omega_p^2}{2 \omega^2 - \omega_p^2}
\right\} = 0 \, ,
\label{twelve}
\end{equation}
where the inclusion of a gap  is described by

\begin{equation}
C= \frac{2\mu}{\pi\hbar^2}\left\{ 1-\frac{\Delta^2}{\mu^2} \right\} \ ,
\end{equation}
  $\mu$ is the chemical potential and $\Delta$ is the gap
between valence and conduction bands.
Consequently, Eq.\ (\ref{twelve})  yields the plasmon frequency as follows \cite{NJMH}:

\begin{equation}
\omega^2= K_1 \pm \sqrt{K_2} \ ,
\end{equation}
with $K_1$ and $ K_2$ defined by

\begin{eqnarray}
&&  K_1 = \frac{\pi e^2  C}{\epsilon_s} q_{\parallel}
+\left(\frac{\omega_p}{2} \right)^2
\nonumber\\
&& K_2 = \frac{\pi e^2  C \, \omega_p^2}{\epsilon_s}
 e^{-2 a q_{\parallel}}  q_{\parallel} +\left[
\left(\frac{\omega_p}{2} \right)^2 - \frac{C e^2 \pi }{\epsilon} q_{\parallel} \right]^2 \, .
\end{eqnarray}
In the low wave number limit $q_\parallel \ll 1/a$ these expressions are reduced to:

\begin{eqnarray}
&& \omega_1 \approx 2 e \sqrt{\frac{\pi a  C }{\epsilon_s}} q_{\parallel}   \\
\nonumber
&& \omega_2 \approx \frac{\omega_p}{\sqrt{2}} + \frac{\sqrt{2} \pi  C e^2 }{\epsilon_s \omega_p} q_{\parallel}
\end{eqnarray}
which are both linear in $q_{||}$, differing from the $q_{||}^{1/2}$-dependence for free-standing
graphene or the 2DEG \cite{Wunsch,pavlo,7,8,9,10}.

\begin{figure}[t]
\centering
\includegraphics[width=0.51\textwidth]{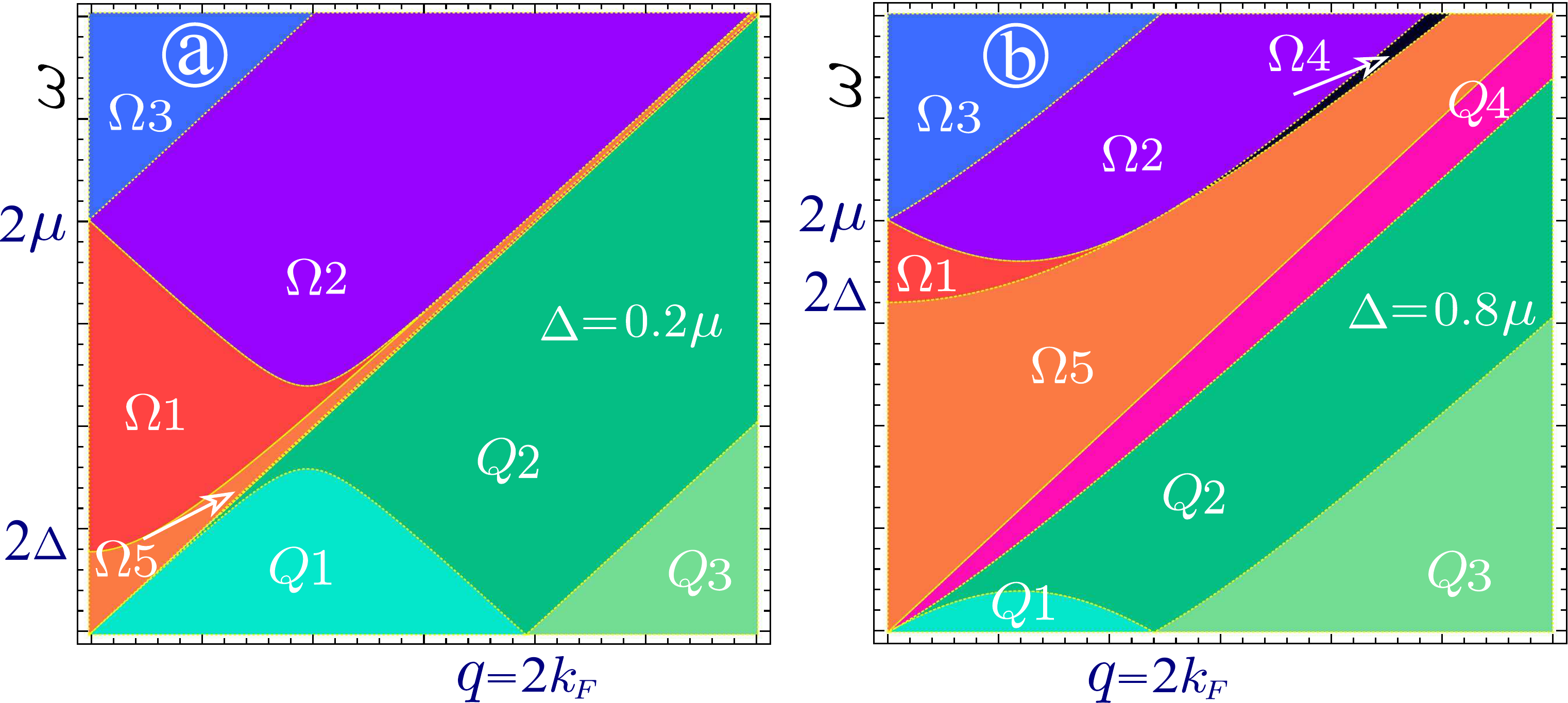}
\caption{(Color online) Schematics showing the regions having differing analytic expressions
 for the non-interaction polarization function $\Pi^{(0)}_{2D} (q_\parallel,\omega) $.
Each part (real or imaginary) is     determined by  a different analytic expression, as in Refs.\
[\onlinecite{pavlo, Silicene, silicene-canada}].
 The regions with $\omega > \hbar v_F q_\parallel$
 are presented as $\Omega 1 - \Omega 5$, while the opposite are given by $ Q 1 - Q 4$.
 Regions with   non-zero $ \Im \Pi^{(0)}_{2D} (q_\parallel,\omega) $ are
 $\Omega_1$, $\Omega_5$, $Q 4$ (where undamped plasmons exist) and $Q 3$. Plot (a) demonstrates the case of a small bandgap  $\Delta = 0.2\,\mu$, whereas the panel $(b)$ shows the case of  relatively large gap $\Delta = 0.6\,\mu$.
}
\label{FIG:5}
\end{figure}

\begin{figure}[ht!]
\centering
\includegraphics[width=0.51\textwidth]{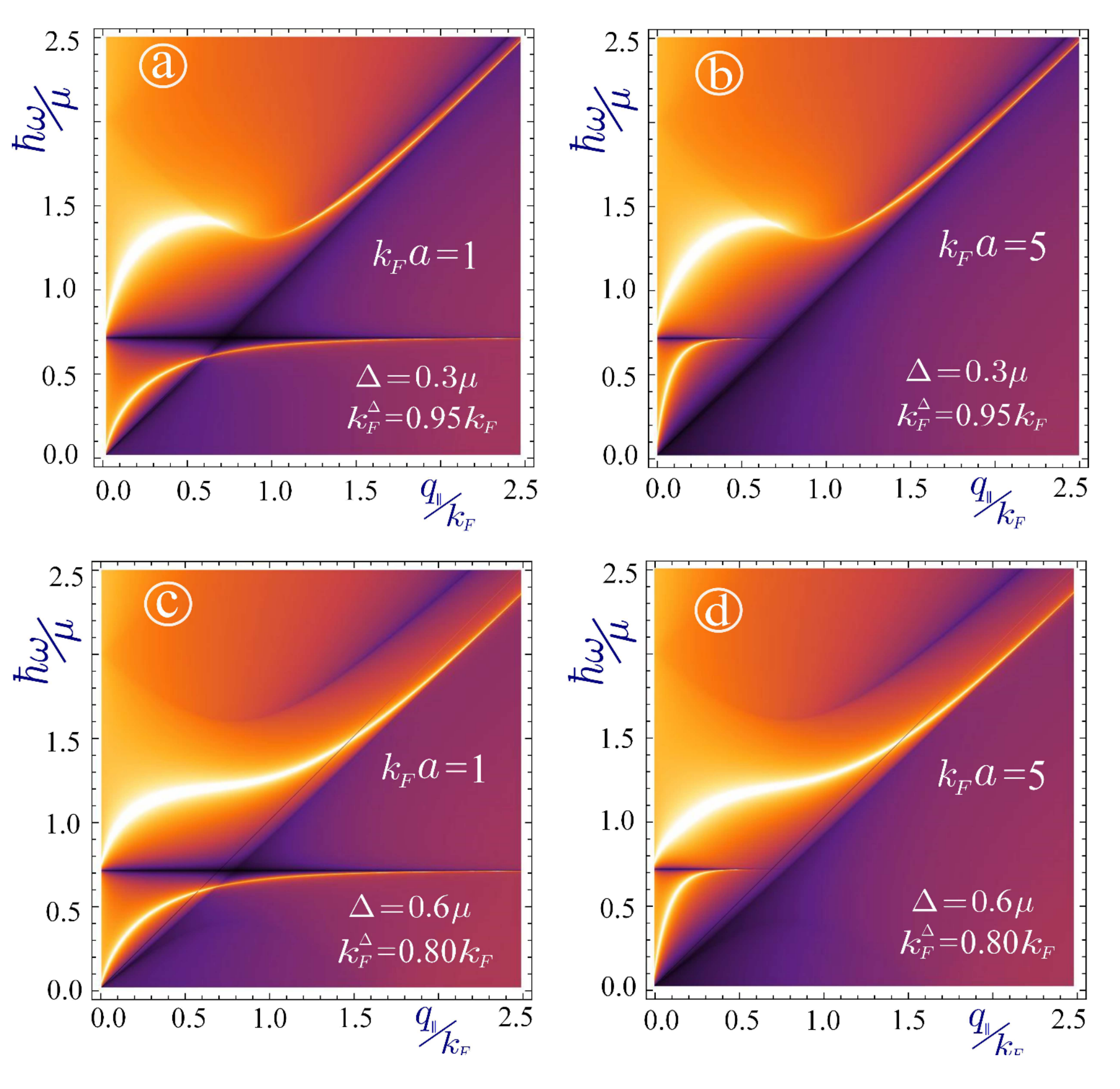}
\caption{(Color online)  Density plot of the real part
of the inverse dispersion function $S_c(q_{\parallel},
\omega+i0^+)$ for extrinsic (doped) graphene having
a finite band gap $(\Delta \neq 0)$ with the peaks corresponding
to the plasmons. The panels are for various values of
the energy gap $\Delta$ and distance $a$ between the surface
and the graphene layer. Panel $(a)$ shows the case when $a =
k_F^{-1} \, \text{and} \, \Delta = 0.3 \mu $, $(b)$ $a =5
k_F^{-1} \, \text{and} \,  \Delta = 0.3 \mu$,  $(c)$  $a =
k_F^{-1} \, \text{and} \, \Delta = 0.6 \mu$ and  $(d)$ describes
$a =5  k_F^{-1} \, \text{and} \, \Delta = 0.6 \mu$. We define
$k_F^\Delta\equiv \sqrt{\mu^2-\Delta^2}/(\hbar v_F)$.}
\label{FIG:45}
\end{figure}

Nonlocality of the graphene plasma introduces changes in the features of
$K(z_1,z_2)$ of Eq. (\ref{eq:GG-inv}) and  in its coupled 2D-surface plasmon
spectrum in two respects.    First, the local coupled mode spectrum  described
in the preceding paragraph is modified by nonlocality corrections in Eq. \
(\ref{eleven}) with the use of the polarization function $\Pi^{(0)}_{2D}(q_\parallel,\omega)$
for all wave numbers as calculated by \cite{pavlo} for gapped graphene.
Secondly, nonlocality introduces  natural damping through the occurrence of
regions in which plasmons can decay  into electron-hole pairs consistent with
energy-momentum conservation.  The intersection of the plasmon dispersion curve $\omega(q_\parallel)$
  with such a particle-hole excitation region (PHER)signals the onset of damping  at T=0K
	with $\Im \ S_{C}(q_{||},\omega) \neq 0$. However, it is the undamped coupled
	plasmons that are of interest with
	
	\begin{equation}
\Im\ S_{C}(q_{||},\omega)=  \Im\ \left(\Pi_{2D}^{(0)} (q_{||},\omega)
\right)\  \frac{2\pi e^2}{\epsilon_s q_{||}}
\left\{ 1+ e^{-2q_{||} a}  \frac{1-\epsilon_B(\omega)}{1+\epsilon_B(\omega)}   \right\} =0\ .
\label{eleven-2}
\end{equation}
The features of the interacting graphene-surface plasmon  spectrum are analyzed here numerically
using the real and imaginary parts of the polarizabilities of Wunsch \cite{Wunsch}  for gapless graphene and
 Pyatkovskiy  \cite{pavlo} for gapped graphene (all at zero temperature).


\par

\section{Calculated Results and discussion}
\label{sec3}

First, we consider graphene with no energy gap and linear energy dispersion for the valence and
conduction bands. The boundaries of the particle-hole modes region are linear, enclosing
 a triangular region, where the plasmons are not damped.  The plasmons for gapless graphene
are shown in Fig.[\ref{FIG:2}]. We discern two plasmon branches,  one attributed to
the surface (the upper branch, originating from $\omega_p/\sqrt{2}$ frequency) and  the other to the
graphene sheet (starting at the origin). We present results for various values of the distance
$a$ between the layer and the surface.
When this   separation is increased, the two branches evolve into a merged spectral line, similar
to the plasmon of extrinsic gapless graphene. The surface plasmon branch tends to be dispersionless
and to exist in the long wave length limit only. For all presented cases, the upper plasmon
mode shows a stronger and broader peak.  We display the absolute value of the real part of
$S_{C}^{-1}(q_{||},\omega)$ to emphasize each peak.


\par

We also solve the   equation $\Re \ S_C (q_\parallel, \omega) = 0$ numerically,
demonstrating the exact solution for the plasmon dispersion relation for both cases of zero
(see Fig.\ \ref{FIG:3}) and finite (Fig.\ \ref{FIG:4}) energy  band gap. These solutions become
 extremely interesting  when the upper branch splits into two parts for the case of small
energy gap.  When the gap is  zero, once again we see that the upper branch (which we
attribute to the presence of a surface) adopts certain features of the plasmon in gapless
graphene mainly because the branch is located in the same $\{ \omega, q_\parallel \}$ regions, both
inside and outside the PHER. However, according to our analytical  results, for long wavelengths
both branches possess finite slope, in contrast to $\backsim \sqrt{q_\parallel}$ behavior in free standing
graphene.

\begin{figure}[ht!]
\centering
\includegraphics[width=0.51\textwidth]{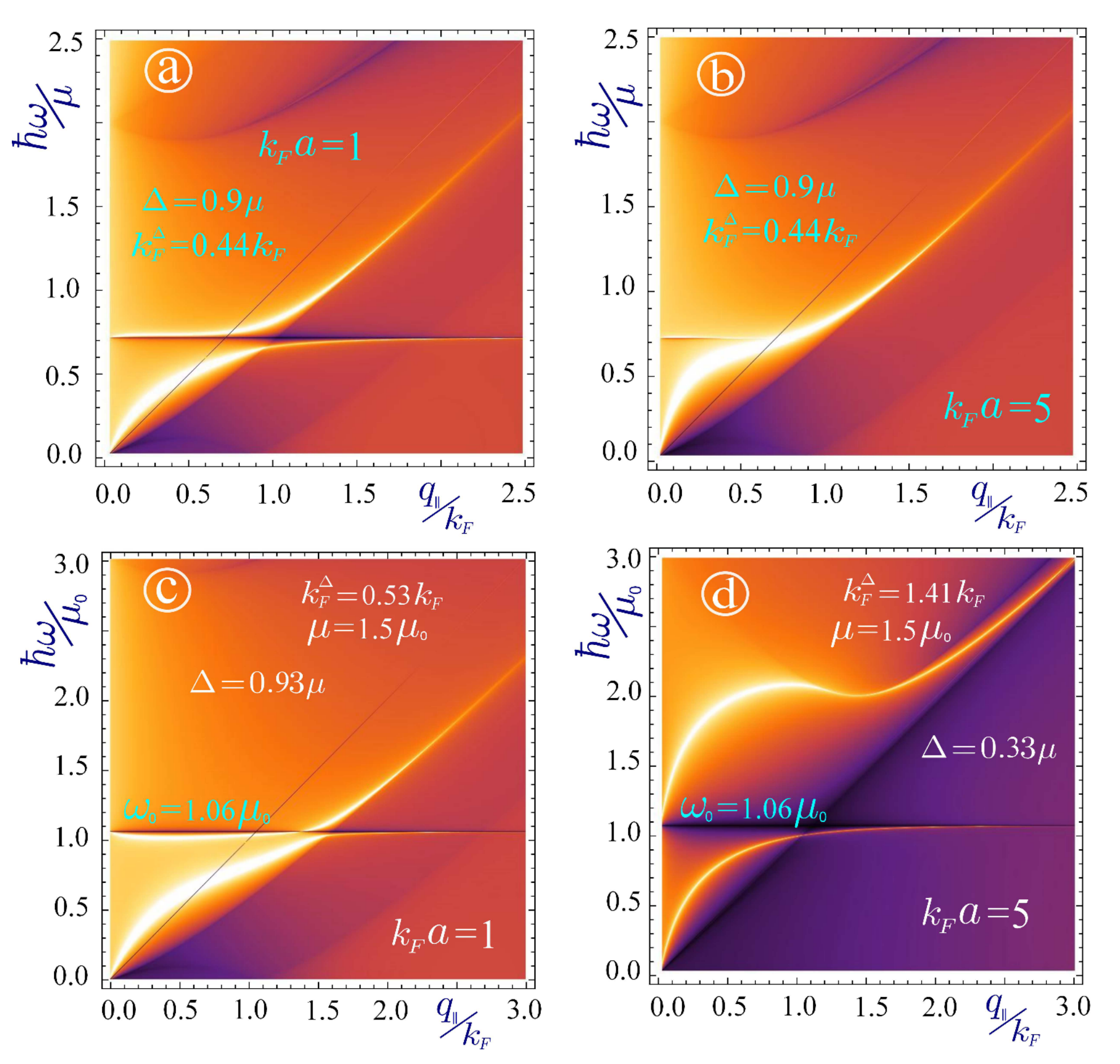}
\caption{(Color online)  Density plot of the real part of  the inverse dispersion function
$S_c(q_{\parallel}, \omega)$ for extrinsic (doped) graphene 
 when the band gap is finite and where  its peak positions
correspond to the plasmon frequencies.
The results in the panels were obtained for various  chosen values of the energy gap
$\Delta$, the distance $a$ between the surface and the graphene layer and the chemical
potential $\mu$, so that panel $(a)$ shows the case when   $ k_F a = 1 \, \text{and} \, \Delta = 0.6 \mu $,
$(b)$ \ $ k_Fa =5   \, \text{and} \,  \Delta = 0.6 \mu$,  $(c)$\  $ \mu= 1.5 \mu_0, \,  k_Fa =1 \,
\text{and} \, \Delta = 0.93 \mu$ and  in $(d)$   $\mu= 1.5 \mu_0, \,  k_Fa = 1\, \text{and} \,
\Delta = 0.33 \mu$. In our notation,   $\mu_0$ is an arbitrary doping, parameter in terms of
which we measure chemical potential. We introduced $k_F^\Delta\equiv \sqrt{\mu^2-\Delta^2}/(\hbar v_F)$. }
\label{FIG:7}
\end{figure}

\medskip
\par

The case of a  small energy gap is presented in Fig.\ \ref{FIG:4} for various energy gap 
and doping values. Similar to free standing graphene, the upper branch is extended due to splitting of
the PHER. It might  also  be split into two different branches as mentioned in  Ref.\  [\onlinecite{pavlo}].
When the distance $a$ of the 2D layer from the surface  is increased, the two plasmon branches
merge into a single branch, which is similar to the plasmon dispersion in gapped graphene.
The general conclusion is that when one of the factors (energy gap, chemical potential or
the separation $a$) is appreciable, the changes caused by a sizable change in one of
the others is not significant.
\par

\par

The role played by the energy band gap is an important part of our investigation. For
 monolayer graphene, an energy  gap leads to to an extended region of undamped plasmons
 \cite{pavlo}. In Fig.\ \ref{FIG:5}, we present the regions of the real and imaginary
parts of the   non-interacting polarization function which have distinct functional forms. We pay
particular attention to the regions outside of the single-particle excitation continuum
since, as   mentioned previously, they encompass   plasmon frequencies
in the domains of $\{ \omega, q_\parallel \}$   where the plasmons are not damped. We denote
these planar regions ($\Omega1$, $\Omega5$ and $Q4$)    with reddish colors. The
condition $\Im \Pi^{(0)}_{2D} (q_\parallel, \omega) = 0$ is   also satisfied in  $Q3$, but
no plasmons are observed in this region. Region $Q4$ with $\hbar v_F q_\parallel > \omega$
plays a crucial role in our study because   this is where the extended undamped
lower plasmon branch is located. This is a   new situation, which was not
encountered in previous works of Refs.\ [\onlinecite{Wunsch, pavlo, 7+, SDSLi}] and it is attributed
to  screening by the carriers in the thick substrate adjoining the 2D layer.

\par
\medskip

 Figure\ \ref{FIG:7}  exhibits our results for plasmon excitations of
a composite system consisting of a layer of gapped graphene   and a thick substrate
for  various values of the energy gap, chemical potential and the distance between the
two bodies. The PHER and its boundaries constitute an important factor determining the plasmons.
Consequently, the upper branch, located mainly in $\Omega1$ and $\Omega5$ regions,  bears
some similarity to the plasmons in  free standing  gapped graphene, including its
splitting into two parts in the vicinity of the boundary of $\Omega_2$. The results
for both  the lower and upper  branches definitely depend on the gap. In the long wavelength
limit, we demonstrate that $\omega_1 \backsimeq \sqrt{C}$ and $\omega_2 \backsimeq \omega_p/\sqrt{2} +
\cdots \backsimeq C$, where $ C \backsimeq \left( 1 - \Delta^2/\mu^2 \right)$.  The plasmon
dispersion relation for  a free standing graphene layer with a finite energy gap is
$\omega \backsimeq \sqrt{C q_\parallel}$, which   differs from our solution and Ref.\  [\onlinecite{NJMH}].
However there is an interesting similarity in that the plasmon frequency is decreased
with increased energy gap. This dependence is observed for increased  values of $q_\parallel$.

\par

The important differences in the plasmon   spectra between free standing graphene and
graphene interacting with a half space arise   from the lower plasmon branch which lies on
both sides of the straight line  $\omega = v_F q_\parallel$ and has a linear dispersion
for small $q_\parallel$. According to  previously published results \cite{pavlo}, the
size of the $Q4$ region is determined by  doping as well as the energy gap. The boundary between
$Q4$ and $Q2$ (with   finite $\Im \Pi^{(0)}_{2D}(q_\parallel,\omega)$) is described
by  $\omega = - \mu + \sqrt{(\hbar v_F)^2 (q_\parallel + k_F)^2 + \Delta^2}$ with $\hbar v_F k_F =\sqrt{\mu^2 -
\Delta^2}$. For $\Delta = 0$, this boundary line is reduced to
$\omega =  v_F q_\parallel $. The plasmon dispersion for various doping concentrations is
presented in Fig.\  \ref{FIG:7}.  Increasing both $\mu$ and $\Delta$, we find more
extended branches where undamped plasmons exist. Figure\ \ref{FIG:7}(d)  clearly demonstrates  
\textit{anti-crossing} and an extended region of undamped plasmons for both branches. In all
cases, the lower plasmon branch does not rise above the line $\omega = \omega_p / \sqrt{2}$.
The curvature of the upper branch is determined by the ratio $\Delta/\mu$ rather rather than
by the gap itself. For certain values of this ratio, the upper branch consists of two different,
separated plasmon branches.
\par

    We note that the exact numerical solutions   in Fig.\ \ref{FIG:4} corresponding to
	$\Re\ S_C(q_\parallel,\omega)=0$  are in agreement with the data for the
		density plots in Figs.\  \ref{FIG:45} and \ref{FIG:7}.
		The results  in these plots confirm the anti-crossing and the extension of
the lower plasmon branches with   increased doping and energy gap. We also note
  that for   large values of the ratio $\Delta / \mu \geq 0.9$ the lower branch becomes
  nearly dispersionless.



\section{Concluding Remarks}
\label{sec4}

In summary, we have calculated the nonlocal plasmon dispersions within  RPA for monolayer graphene interacting with
a substrate, for arbitrary wavelength. In this, we  investigated numerically the effects of the energy gap
for extrinsic  graphene, as well  as  the effects of its distance from the surface, on the plasmon dispersion
relation.  Our considerations were motivated by  recent experimental work showing a linear plasmon
dispersion in the long wavelength limit \cite{1+} and the subsequent theoretical work by one of
the authors \cite{NJMH} to account for this observation, which is extended here to a fully general numerical
description of nonlocal effects in monolayer graphene when the separation $a$ is varied and when the energy
gap is increased.  Our new results in this paper vividly demonstrate  that a thorough
investigation necessitates  incorporating the polarization into the dispersion  equation
at shorter wavelengths.

\par

The distance $a$ between monolayer  graphene and the surface was  varied in our nonlocal numerical
calculations.   In all cases, there are two plasmon branches; one originating from the surface
plasmon and the other from the graphene layer. Both gapless and  gapped graphene have been investigated.
The most important consequence of introducing the energy gap in graphene is the extended
region of undamped plasmons for both branches. Specifically, referring to Fig. \
 \ref{FIG:3}(a), we note that   the upper plasmon dispersion  curve enters the gap in
 the particle-hole spectrum like that for gapped free standing graphene and these two
 curves are close to each other within this gap.
In addition,  the lower plasmon branch is   undamped for a wider range of   wave vectors
$q_\parallel$ by entering the gap in the particle-hole region. As revealed in Fig.\
\ref{FIG:4}(c), the lower branch may anti-cross with the upper one   for sufficiently
high doping concentration and large band gap.    Both plasmon frequencies decrease with
increased energy gap. This is also the behavior  for free standing  gapped graphene.
However, the exact mathematical dependence is different in each case.
Also, either one of the plasmon branches may  bifurcate into two branches in the
the single-particle excitation region, as demonstrated in Fig.\ \ref{FIG:7}(b).
These new results for the plasmons may potentially lead to a number of
applications in electronic devices since the plasmons play an important role in the response
properties to external electromagnetic fields.

\acknowledgments
This research was supported by  contract \# FA 9453-13-1-0291 of
AFRL. We thank Danhong Huang for helpful discussions.


\end{document}